\documentclass[12pt]{article}
\usepackage{latexsym,amsmath,amsfonts,amssymb}
\newtheorem{Theorem}{Theorem}
\author{Renat Zhdanov\thanks{E-mail:\ renat.zhdanov@bio-key.com}\\
BIO-key International, Eagan, MN, USA}
\title{Towards classification of quasi-local symmetries of evolution
equations}
\date{}

\begin{document}
\maketitle

\begin{abstract}
We develop efficient group-theoretical approach to the problem of
classification of evolution equations that admit non-local
transformation groups (quasi-local symmetries), i.e., groups
involving integrals of the dependent variable. We classify
realizations of two- and three-dimensional Lie algebras leading to
equations admitting quasi-local symmetries. Finally, we generalize
the approach in question for the case of an arbitrary system of
evolution equations with two independent variables.
\end{abstract}

\section{Introduction.}

Consider the general evolution equation in one spatial dimension
\begin{equation}
\label{r1}
u_t=F(t,x,u,u_1,u_2,\ldots,u_n),\quad n\ge2,
\end{equation}
where $u=u(t,x)$ is a real-valued function of two real variables $t,
x$, $u_i={\partial^i u}/{\partial x^i},\ i=1,2,\ldots, n$, and $F$
is an arbitrary smooth real-valued function.

The most general Lie transformation group leaving differential
equation (\ref{r1}) invariant has the form (see, e.g.,
\cite{ren1,ren2})
\begin{equation}
\label{r2}
t'=T(t,{\vec{\theta}}\,),\quad x'=X(t,x,u,{\vec{\theta}}\,),\quad
u=U(t,x,u,{\vec{\theta}}\,).
\end{equation}
Here $T,X,U$ are smooth real-valued functions satisfying the
non-singularity condition $\frac{D(T,X,U)}{D(t,x,u)} \not\equiv 0$
in some open domain of ${\mathbb{R}}^3$ and $\vec \theta =
(\theta_1,\theta_2,\ldots,\theta_r)\in {\mathbb R}^r$ is the vector
of group parameters.

If a transformation of the space of variables $t,x,u$ changes the
specific form of (\ref{r1}) leaving invariant its differential
structure, then we arrive at the concept of {\it equivalence\
group}. More precisely, if the (locally) invertible change of
variables,
$$
t\to t'=T(t,x,u),\quad x\to x'=X(t,x,u),\quad u\to u'=U(t,x,u)
$$
maps Eq.(\ref{r1}) into a possibly different $n$-th order evolution
equation
$$
u'_{t'}=G(t',x',u',u'_1,u'_2,\ldots,u'_n),
$$
then this change of variables is called equivalence transformation.
The set of all possible equivalence transformations forms a
diffeomorphism group, ${\cal E}$, and is called the equivalence
group of Eq.(\ref{r1}). Clearly, if we require that $G\equiv F$ then
the equivalence group reduces to the symmetry group of
Eq.(\ref{r1}). Consequently, Lie symmetry group of a given equation
is a subgroup of its equivalence group.

Let partial differential equation (\ref{r1}) be invariant under Lie
transformation group (\ref{r2}). What would happen to this Lie
symmetry if we perform a transformation from the equivalence group
of the equation under study? Evidently, transformation group
(\ref{r2}) after being rewritten in the 'new' variables $t',x',u'$
becomes Lie symmetry of the transformed equation.

Now suppose that we allow for the more general equivalence
transformation group
$$
t\to t'=T(t,x,u,\vec v),\quad x\to x'=X(t,x,u,\vec v),\quad u\to
u'=U(t,x,u, \vec v),
$$
where $\vec v=(u_1,u_2,\ldots,u_p, \partial^{-1}u, \partial^{-2}u,
\ldots\partial^{-s}u)$ with $\partial^{-1}u=\int u(t,x)dx$ and
$\partial^{-k-1}=\partial^{-1}\partial^{-k}$. Saying it another way,
we allow for an equivalence transformation to include derivatives
and integrals of the dependent variable $u$. If such a
transformation still preserves the differential structure of
evolution equation (\ref{r1}), what would happen to Lie symmetries
of the latter? The answer is, 'it depends'. In some cases, Lie
symmetry transforms into another Lie symmetry. However, it could
happen that some Lie symmetries would 'disappear' after performing
equivalence transformation, meaning that they cannot be found within
the framework of the infinitesimal Lie method. The reason is that
the transformation rule for the variables $t,x,u$ might contain
derivatives and integrals of $u$, which are beyond reach of the
standard Lie method. One needs to apply the generalized Lie
\cite{ibr1}-\cite{blu1} or non-Lie \cite{fus1} approaches to be able
to handle those symmetries.

To single out Lie symmetries, which after non-local equivalence
transformation of an equation under study turn into non-Lie
symmetries Ibragimov et al \cite{ibr2} introduced the term 'quasi-local
symmetry', which we use throughout the paper. Independently, the
concept of quasi-local symmetry has been suggested in \cite{puk}.

In the present paper we suggest a simple regular method for deriving
quasi-local symmetries (QLS) of evolution equations. Note that the
basic idea of the method has been suggested in our paper \cite{ren3}
and some non-trivial examples of second-order evolution equations
with QLS are given in \cite{ren2}.

Next, we demonstrate how to apply the method developed to arbitrary
systems of evolution equations with two independent variables.

\section{Method description.}

The most general Lie transformation group admitted by evolution
equation (\ref{r1}) is of the form (\ref{r2}). The infinitesimal
operator, $Q$, of this group reads as \cite{ren1}
\begin{equation}
\label{r3}
Q=\tau(t)\partial+\xi(t,x,u)\partial_x+\eta(t,x,u)\partial_u.
\end{equation}
Provided, $\tau\equiv 0$ there is a transformation,
\begin{equation}
\label{r100}
t\to \bar t=t,\quad x\to \bar x=X(t,x,u),\quad u\to \bar u=U(t,x,u),
\end{equation}
that reduces $Q$ to the canonical form $\partial_u$ (we drop the
bars). Evolution equation (\ref{r1}) now becomes
\begin{equation}
\label{r4}
u_t=f(t,x,u_1,u_2,\ldots,u_n).
\end{equation}
Note that the right-hand side of Eq.(\ref{r4}) does not depend
explicitly on $u$.

Differentiating (\ref{r4}) with respect to $x$ yields
$$
u_{tx}=\frac{\partial f}{\partial x} +
\sum_{i=1}^{n}\,\frac{\partial f}{\partial u_i}u_{i+1}.
$$
Making the change of variables
\begin{equation}
\label{r5}
\bar t=t,\quad \bar x=x,\quad \bar u=u_x
\end{equation}
and dropping the bars we finally get
\begin{equation}
\label{r6}
u_{t}=\frac{\partial f}{\partial x} + \frac{\partial f}{\partial u}
+ \sum_{i=2}^{n}\,\frac{\partial f}{\partial u_{i-1}}u_{i},
\end{equation}
where $f=f(t,x,u,u_1,\ldots,u_{n-1})$.

Thus non-local transformation (\ref{r5}) preserves the differential
structure of the class of evolution equations (\ref{r4}).

Let differential equation (\ref{r4}) admit $r$-parameter Lie
transformation group (\ref{r2}) with $\vec \theta =
(\theta_1,\ldots,\theta_r)$ and $r\ge 2$. To obtain the symmetry
group of Eq.(\ref{r4}) we need to transform (\ref{r2}) according to
(\ref{r5}). To this end we compute the first prolongation of
formulas (\ref{r2}) and derive the transformation rule for the first
derivative of $u$
$$
\frac{\partial u'}{\partial x'}=\frac{U_uu_x+U_x}{X_uu_x+X_x}.
$$
So symmetry group (\ref{r2}) now reads as
\begin{equation}
\label{r7}
t'=T(t,\vec\theta\,),\quad x'=X(t,x,v,\vec\theta\,),\quad
u'=\frac{U_vu+U_x}{X_vu+X_x}
\end{equation}
with $v=\partial^{-1}u$ and $U=(t,x,v,\vec\theta\,)$. Consequently,
if the right-hand sides of (\ref{r7}) depend explicitly on the
non-local variable $v$, then transformation group (\ref{r7}) is a
quasi-local symmetry of Eq.(\ref{r6}).

Evidently, transformations (\ref{r7}) include variable $v$
if and only if
$$
X_v \ne 0\ {\rm or}\ \frac{\partial}{\partial
v}\left(\frac{U_vu+U_x}{X_vu+X_x}\right)\ne 0
$$
or, equivalently (since all the functions involved are real-valued),
\begin{eqnarray}
&&(X_v)^2 + (U_{vv}X_v-U_vX_{vv})^2 + (U_{xv}X_x-U_xX_{xv})^2
\nonumber\\
&&\quad + (U_{vv}X_x+U_{xv}X_v - U_xX_{vv} - U_vX_{xv})^2 \ne
0.\label{r8}
\end{eqnarray}
If $X_v\not=0$, then the above inequality holds true. If $X_v$ does
vanish identically, then (\ref{r8}) reduces to
$U_{xv}^2+U_{vv}^2\not=0$. It is straightforward to express the
above constraints in terms of the coefficients of the corresponding
infinitesimal operator of group (\ref{r2}). As a result, we get the
following assertion.
\begin{Theorem}
Equation (\ref{r4}) can be reduced to evolution equation (\ref{r6})
having QLS if it admits Lie symmetry, whose infinitesimal generator
satisfies one of the inequalities
\begin{eqnarray}
&&\frac{\partial \xi}{\partial u} \not=0, \label{r9}\\
&&\frac{\partial\xi}{\partial u}=0,\quad
\left(\frac{\partial^2\eta}{\partial u\partial x}\right)^2 +
\left(\frac{\partial^2\eta}{\partial u\partial u}\right)^2 \not=
0.\label{r10}
\end{eqnarray}
\end{Theorem}

Now we can formulate an algorithm for constructing evolution
equations of the form (\ref{r1}) admitting QLS.
\begin{enumerate}
\item We compute the maximal Lie symmetry group ${\cal S}$ of
differential equation (\ref{r1}).
\item We classify inequivalent one-parameter subgroups of ${\cal
S}$ and select subgroups ${\cal S}_1,\ldots{\cal S}_p$ whose
infinitesimal operators are of the form $Q=\xi(t,x,u)\partial_x +
\eta(t,x,u)\partial_u$.
\item For each subgroup, ${\cal S}_i$, we construct a change of
variables (\ref{r100}) reducing the corresponding infinitesimal
operator $Q$ to the canonical operator $\partial_{\bar u}$, which
leads to evolution equations of the form (\ref{r4}).
\item Since the invariance group, $\bar{\cal S}$, admitted by
(\ref{r4}) is isomorphic to ${\cal S}$, we can utilize the results
of subgroup classification of ${\cal S}$. For each of the
one-parameter subgroups of $\bar{\cal S}$ we check whether its
infinitesimal generators satisfies one of conditions (\ref{r9}),
(\ref{r10}) of Theorem 1. This yields the list of evolution
equations that can be reduced to those having QLS.
\item Performing non-local change of variables (\ref{r5}) yields
evolution equations (\ref{r6}) admitting quasi-local symmetries
(\ref{r7}).
\end{enumerate}

Full implementation of the above approach will be the topic of our
future publication. Here we restrict our considerations to
classifying realizations of two- and three-parameter Lie
transformation groups leading to evolution equations (\ref{r1}) that
admit QLS.

Hereafter we suggest that evolution equation (\ref{r1}) admits a Lie
symmetry $Q=\xi(t,x,u)\partial_x + \eta(t,x,u)\partial_u$ and
therefore can be reduced to the form (\ref{r4}). Differential
equation (\ref{r4}) is guaranteed to admit at least a one-parameter
symmetry group, which is generated by the operator $\partial_u$.
What we are going to do is to describe all realizations of two- and
three-dimensional Lie algebras, which
\begin{itemize}
\item are Lie symmetry algebras of equations of the form (\ref{r4}),
and,
\item have coefficients satisfying one of the inequalities
(\ref{r9}), (\ref{r10}) from Theorem 1.
\end{itemize}

With these realizations in hand, the problem of describing equations
having QLS reduces to a straightforward application of the
infinitesimal Lie method \cite{olv1,ovs1,fus2}, which boils down to
integrating Euler-Lagrange system for calculating differential
invariants of the corresponding Lie algebras of first-order
differential operators.

Let us remind that the most general symmetry generator admitted by
(\ref{r4}) is of the form (\ref{r3}), while the most general
equivalence group admitted by Eq.(\ref{r1}) reads as (see, e.g.,
\cite{kin1})
\begin{equation}
\label{r11}
\bar t = T(t),\quad \bar x = X(t,x,u),\quad \bar u = U(t,x,u),
\end{equation}
where $T,X,U$ are arbitrary smooth real-valued functions.

We can always choose basis operators of Lie symmetry algebra of
Eq.(\ref{r4}) so that
$$
e_0=\partial_u,\quad e_i=\tau_i(t)\partial_t+\xi_i(t,x,u)\partial_x+
\eta_i(t,x,u)\partial_u,
$$
where $i=1,\ldots,r$. Note that by the Magadeev theorem \cite{mag1}
the maximal possible value for $r$ is $n+4$, $n$ being the order of
evolution equation (\ref{r4}), provided (\ref{r4}) is not locally
equivalent to a linear equation. In particular, for the second-order
evolution equation we have $r<=6$. By the definition of Lie algebra
there are constant $r\times r$ matrix $C$ and constant $r$-component
vector $\vec c$ such that
\begin{equation}
\label{r12}
[e_0,\, e_i]=\sum_{j=1}^r\,C_{ij}e_j+c_ie_0,\quad i=1,\ldots,r.
\end{equation}
Here $[Q_1,\,Q_2]\equiv Q_1Q_2-Q_2Q_1$.

System of equations (\ref{r12}) is the starting point of our
classification algorithm. First of all, let us note that by
re-arranging the basis of the Lie algebra, $e_{\mu}\to
\sum_{\nu=0}^r\,a_{\mu\nu}e_{\nu},\ \mu=0,1,\ldots,r$, we can always
reduce the constant matrix $C$ to the canonical form. Consequently,
without any loss of generality we may assume that the matrix $C$ is
in the canonical form.

Computing commutators in the left-hand sides of (\ref{r12}) and
equating the coefficients of linearly-independent operators
$\partial_t,\partial_x,\partial_u$ we get the following system of
partial differential equations:
\begin{equation}
\label{r13}
\frac{\partial\vec\xi}{\partial u} = C\vec\xi,\quad
\frac{\vec\eta}{\partial u} = C\vec\eta + \vec c,\quad
C\vec\tau = 0,
\end{equation}
where $\vec\xi=(\xi_1,\ldots,\xi_r)$,
$\vec\eta=(\eta_1,\ldots,\eta_r)$,
$\vec\tau=(\tau_1,\ldots,\tau_r)$. After integrating differential
equations (\ref{r13}) we need to ensure that the operators
$e_1,\ldots, e_r$ do form a basis of Lie algebra and satisfy the
additional set of commutation relations,
$$
[e_i,\, e_j]=\sum_{k=1}^r\,c_{ij}^ke_k,\quad k=1,\ldots,r.
$$
Next, we simplify the form of operators $e_1,\ldots,e_r$ using the
suitable equivalence transformations from the group ${\cal E}$. As a
final step, we verify that at least one of the coefficients of one
the operators $e_1,\ldots, e_r$ satisfy either (\ref{r9}) or
(\ref{r10}).

\section {Realizations of two-dimensional QLS algebras}

For the case when $r=1$, system (\ref{r13}) reduces to a pair of
non-coupled differential equations
\begin{equation}
\label{r14}
\xi_u = \lambda\xi,\quad \eta_u = \lambda\eta + c,\quad
\lambda\tau=0,
\end{equation}
where $\lambda, c$ are constants.

While integrating (\ref{r14}) we need to differentiate between the
two cases $\lambda\ne 0$ and $\lambda=0$.

\noindent
{\bf Case 1.}\ $\lambda\ne 0$.
The general solution of (\ref{r14}) has the form
\begin{eqnarray*}
&&\tau(t)=0,\quad \xi(t,x,u)=W_1(t,x)\exp(\lambda u),\\
&&\eta(t,x,u)=W_2(t,x)\exp(\lambda u) - c\lambda^{-1}.
\end{eqnarray*}
where $W_1, W_2$ are arbitrary smooth functions. So that the
two-dimensional Lie algebra $\langle e_0,e_1\rangle$ read as
$$
e_0=\partial_u,\quad e_1=W_1(t,x)\exp(\lambda u)\partial_x+
\left(W_2(t,x)\exp(\lambda u)-c\lambda^{-1}\right)\partial_u.
$$
As $\lambda \ne 0$ we can always re-scale $u$, i.e., make a
transformation $u\to ku,\ k = {\rm const}$, in order to get $\lambda
= 1$. Next taking as new $e_1$ the linear combination
$c\lambda^{-1}e_0+e_1$ we get rid of the term in $e_1$ which is
proportional to $c$, namely,
$$
e_0=\partial_u,\quad
e_1=W_1(t,x)\exp(u)\partial_x+W_2(t,x)\exp(u)\partial_u.
$$
It is not difficult to verify that the most general subgroup of the
equivalence group ${\cal E}$ not altering the form of equation
(\ref{r4}) and operator $\partial_u$ is given by the formulas
\begin{equation}
\label{r15}
\bar t = T(t),\quad \bar x = X(t,x),\quad \bar u = u + U(t,x),
\end{equation}
where $T,X,U$ are arbitrary smooth functions.

Since the functions $W_1$ and $W_2$ do not vanish simultaneously, we
have three possible subcases, (1) $W_1\ne 0, W_2\ne 0$, (2)
$W_1\ne0, W_2=0$, and $W_1=0, W_2\ne0$.

\noindent
{\bf Case 1.1.}\ $W_1\ne 0, W_2\ne 0$.
Applying (\ref{r15}) with $T=t$ we reduce the operator $e_1$ to the
form
$$
e_1=\epsilon_1\exp(u)\partial_x+\epsilon_2\exp(u)\partial_u,
$$
where $\epsilon_1=\pm 1, \epsilon_2=\pm 1$. Combining equivalence
transformation $t\to t,\ x\to -x,\ u\to u$ and re-scaling $e_1\to
-e_1$ we get the final form of the basis elements $e_0, e_1$
$$
e_0=\partial_u,\quad e_1=\exp(u)(\partial_x+\partial_u).
$$
Since $\xi_u=\exp(u)\ne 0$, the condition (\ref{r9}) of Theorem 1
holds true and the evolution equation invariant under the above
algebra is equivalent to a quasi-linear evolution equation that
admits QLS.

\noindent {\bf Case 1.2.}\ $W_1\ne 0, W_2\ne 0$. Applying
transformation (\ref{r14}) with $T(t)=t, U(t,x)=0, k=1$ we reduce
the operator $e_1$ to the form $ e_1=\exp(u)\partial_x. $ Again, the
coefficient $\xi=\exp(u)$ obeys condition (\ref{r9}) of Theorem 1
and, therefore, it gives rise to the two-dimensional Lie algebra
$$
e_0=\partial_u,\quad e_1=\exp(u)\partial_x
$$
that leads to an evolution equation admitting QLS.

\noindent {\bf Case 1.3.}\ $W_1=0, W_2\ne 0$. Applying
transformation (\ref{r14}) with $T(t)=t, X(t,x)=x$ we reduce the
operator $e_1$ to the form $ e_1=\pm\exp(u)\partial_u$. Re-scaling,
if necessary, the operator $e_1$ to $-e_1$ we can make sure that
$e_1$ reads as $\exp(u)\partial_u$ and finally get
$$
e_0=\partial_u,\quad e_2=\exp(u)\partial_u.
$$
Since the coefficient $\eta=\exp(u)$ obeys condition (\ref{r10}) of
Theorem 1, an evolution equation invariant under the above algebra
is equivalent to a partial differential equation of the form
(\ref{r6}) which has QLS.

\noindent {\bf Case 2.}\ $\lambda=0$. System (\ref{r14}) is readily
integrated to yield
$$
T(t)=W_0(t),\quad \xi=W_1(t,x),\quad \eta=W_2(t,x)+cu.
$$
Here $W_0, W_1, W_2$ are arbitrary smooth functions. Checking the
conditions of Theorem 1 we see that neither of them can be satisfied
by the coefficients of the operator $e_1$. Consequently, this case
yields no equations admitting QLS.

Below we give the full list of ${\cal E}$-inequivalent realizations
of two-dimensional Lie algebras spanned by the operators
$e_0=\partial_u$, $e_1=T(t)\partial_t$ $+\xi(t,x,u)\partial_x$
$+\eta(t,x,u)\partial_u$ satisfying the conditions of Theorem 1
\begin{eqnarray*}
A_{2}^1&:&\langle \partial_u,\
\exp(u)(\partial_x+\partial_u)\rangle,\\
A_{2}^2&:&\langle \partial_u,\ \exp(u)\partial_x\rangle,\\
A_{2}^3&:&\langle \partial_u,\ \exp(u)\partial_u\rangle.
\end{eqnarray*}

Evolution equations (\ref{r4}) invariant under the above algebras
are reduced to differential equations that admit QES. The
corresponding QLS are obtained by re-writing the transformation
groups generated by the operators $e_1$ in terms of new (non-local)
variables $t,x,u$ and $\partial^{-1}u$.

Consider, for example, the algebra $A_{2}^2=\langle \partial_u,\
\exp(u)\partial_x\rangle$. Applying the standard infinitesimal Lie
algorithm \cite{ovs1} we obtain the determining equations for the
function $f$
$$
-u_xf-f_x+u_x^2f_{u_x}+(u_x^3+3u_xu_{xx})f_{u_{xx}}=0.
$$
The general solution of the above equation reads as
$$
f(t,x,u_x,u_{xx})=u_x\tilde f(\omega_0,\omega_1, \omega_2),
$$
where $\tilde f$ is an arbitrary smooth function and $\omega_0=t$,
$\omega_1=(xu_x-1)u_x^{-1}$, $\omega_2=(u_{xx}+u_x^2)u_x^{-3}$ are
absolute invariants of the transformation group generated by the
operators $\partial_u$ and $\exp(u)\partial_x$. Consequently, the
evolution equation invariant under the algebra $A_{2}^2$ is of the
form
$$
u_t=u_x\tilde f(\omega_0,\omega_1, \omega_2).
$$
Differentiating the above equation with respect to $x$ and replacing
$u_x$ with $u$ according to (\ref{r5}) we arrive at the evolution
equation
$$
u_t=u_x\tilde f + \frac{u_x+u^2}{u^2}\tilde f_{\omega_1} +
\frac{uu_{xx}-3(u^2+1)u_x}{u^4}\tilde f_{\omega_2}
$$
with $\omega_0=t$, $\omega_1=x-u^{-1}$, $\omega_2=(u_x+u^2)u^{-3}$.
This differential equation admits the following quasi-local symmetry
group
$$
t'=t,\ x'=x+\theta\exp(v),\quad u'=\frac{u}{1+\theta u\exp(v)}.
$$
where $\theta$ is a group parameter and $v=\partial^{-1}u$.

\section {Realizations of three-dimensional QLS algebras}

Consider system of partial differential equations (\ref{r13}) with
$r=2$. The constant $2\times 2$ matrix $C$ has been reduced to the
canonical real Jordan form. There are three inequivalent cases that
need to be considered separately, namely, when eigenvalues
\begin{enumerate}
\item are complex conjugate,
\item are real, and the matrix $C$ is diagonal matrix,
\item are real and the matrix $C$ is the $2\times2$ canonical
Jordan box
$$
\left(
\begin{array}{cc}
\lambda & 1\\
0 & \lambda
\end{array}
\right).
$$
\end{enumerate}
We consider in detail the class of realizations of three-dimensional
Lie algebras obtained for the case when $C$ has two complex
eigenvalues $\lambda_1, \lambda_2$. For the remaining two classes we
present the final results only.

\subsection {Case of diagonal canonical form with complex
ei\-gen\-values}
As the characteristic equation of the real matrix $C$ is real, the
eigenvalues have to satisfy the additional constraint $\lambda_1^* =
\lambda_2$. Consequently, if we define
$\lambda=(\lambda_1+\lambda_2)/2$ and
$a=(\lambda_1-\lambda_2)/(2i)$, then the general solution of
(\ref{r13}) can be represented in the form
\begin{eqnarray}
&&\tau_1=0,\quad \tau_2=0,\nonumber\\
&&\xi_1=(W_1(t,x)\cos(au)+W_2(t,x)\sin(au))\exp(\lambda
u),\nonumber\\
&&\xi_2=(W_2(t,x)\cos(au)-W_1(t,x)\sin(au))\exp(\lambda
u),\label{r16}\\
&&\eta_1=(W_3(t,x)\cos(au)+W_4(t,x)\sin(au))\exp(\lambda u) +
b_1,\nonumber\\
&&\eta_2=(W_4(t,x)\cos(au)-W_3(t,x)\sin(au))\exp(\lambda u) +
b_2.\nonumber
\end{eqnarray}
Here $W_1, W_2, W_3, W_4$ are arbitrary smooth real-valued
functions, $b_1, b_2$ are real constants. Hence, the most general
form of the basis operators $e_1, e_2$ is
\begin{eqnarray*}
e_1&=&(W_1(t,x)\cos(au)+W_2(t,x)\sin(au))\exp(\lambda u)\partial_x\\
&& + \Bigl((W_3(t,x)\cos(au)+W_4(t,x)\sin(au))\exp(\lambda u) +
b_1\Bigr)\partial_u,\\
e_2&=&(W_2(t,x)\cos(au)-W_1(t,x)\sin(au))\exp(\lambda u)\partial_x\\
&& + \Bigl((W_4(t,x)\cos(au)-W_3(t,x)\sin(au))\exp(\lambda u) +
b_2\Bigr)\partial_u.
\end{eqnarray*}
Note that $a\ne 0$, otherwise, $\lambda_1, \lambda_2$ are not
complex. By re-scaling the variable $u$ we can make $a$ equal to 1.
Next, applying to the operators $e_1, e_2$ an equivalence
transformation (\ref{r14}) with $T(t)=t, X(t,x)=x$ we can get rid of
the function $W_2$. With these remarks the operators $e_1, e_2$ take
the form
\begin{eqnarray}
e_1&=&\Bigl((W_3(t,x)\cos{u}+W_4(t,x)\sin{u})\exp(\lambda u) +
b_1\Bigr)\partial_u\nonumber\\
&& + W_1(t,x)\cos{u}\exp(\lambda u)\partial_x, \label{r17}\\
e_2&=& \Bigl((W_4(t,x)\cos{u}-W_3(t,x)\sin{u})\exp(\lambda u) +
b_2\Bigr)\partial_u\nonumber\\
&&-W_1(t,x)\sin{u}\exp(\lambda u)\partial_x.\nonumber
\end{eqnarray}
In what follows we need to distinguish between the cases,
$\lambda\ne 0$ and $\lambda=0$.

\noindent {\bf Case 1.}\ $\lambda\ne 0$. Performing, if necessary,
transformation (\ref{r14}) with $T(t)=t, U(t,x)=u$ we can always
make non-vanishing identically function $W$ equal to 1. Next, taking
as $e_1$ and $e_2$ the linear combinations $e_1-b_1e_0$ and
$e_2-b_2e_0$ we can get rid of parameters $b_1, b_2$.

Now we need to ensure that the operators $e_0, e_1, e_2$ do form a
realization of a Lie algebra. To this end we have to verify that the
relation
\begin{equation}
\label{r18}
[e_1,\ e_2]=\alpha e_1 + \beta e_2 + \gamma e_0
\end{equation}
with some real $\alpha,\beta,\gamma$ holds true. Calculating the
commutators and equating the coefficients of linearly-independent
operators $\partial_t,\partial_x,\partial_u$ we get the system of
differential equations for $W_3, W_4$. Its general solution is given
by the formulas
$$
W_3=\lambda(\lambda^2+1)^{-1}(x+p(t))^{-1},\quad
W_4=-(\lambda^2+1)^{-1}(x+p(t))^{-1}.
$$
Here $p(t)$ is an arbitrary smooth function. Making the equivalence
transformation (\ref{r14}) with $T(t)=t, X(t,x)=p(t), U(t,x)=0$ we
eliminate the function $p(t)$ and arrive at the following
realization of a three-dimensional Lie algebra
\begin{eqnarray*}
e_0&=&\partial_u,\\
e_1&=&\exp(\lambda u)\cos{u}\partial_x
+(\lambda^2+1)^{-1}(\lambda\cos{u}-\sin{u})x^{-1}\exp(\lambda
u)\partial_u,\\
e_2&=&-\exp(\lambda u)\sin{u}\partial_x
-(\lambda^2+1)^{-1}(\cos{u}-\lambda\sin{u})x^{-1}\exp(\lambda
u)\partial_u,
\end{eqnarray*}
Evidently, the coefficients of $e_1, e_2$ satisfy condition
(\ref{r9}) of Theorem 1 and, consequently, evolution equation
invariant under the symmetry algebra $e_0$, $e_1$, $e_2$ can be
reduced to the one having QLS.

\noindent
{\bf Case 2.}\ $\lambda=0$. Operators (\ref{r17}) take the form
\begin{eqnarray*}
e_1&=&\Bigl((W_3(t,x)\cos{u}+W_4(t,x)\sin{u}) + b_1\Bigr)\partial_u
+ W_1(t,x)\cos{u}\partial_x, \nonumber\\
e_2&=& \Bigl((W_4(t,x)\cos{u}-W_3(t,x)\sin{u}) + b_2\Bigr)\partial_u
-W_1(t,x)\sin{u}\partial_x.\nonumber
\end{eqnarray*}
Replacing $e_1$ and $e_2$ with the linear combinations $e_1-b_1e_0$
and $e_2-b_2e_0$ eliminates the parameters $b_1, b_2$. So that we
can assume that $b_1=0, b_2=0$ without any loss of generality.

\noindent {\bf Case 2.1.}\ $W_1\ne 0$. Utilizing equivalence
transformation (\ref{r14}) with $T=t, U=0$ we can make $W_1$ equal
to 1. After simple algebra we prove that for the operators $e_1,
e_2$ to satisfy the remaining commutation relation (\ref{r18}) the
functions $W_3, W_4$ have to take one of the following forms:
\begin{eqnarray*}
&&W_3=0,\quad W_4=\mu\tan(\mu x),\\
&&W_3=0,\quad W_4=-\mu\tanh(\mu x),\\
&&W_3=0,\quad W_4=x^{-1},
\end{eqnarray*}
where $\mu$ is an arbitrary real parameter. Inserting the above
expressions into the corresponding formulas for $e_1, e_2$ we
finally get
\begin{eqnarray*}
e_0&=&\partial_u,\\
e_1&=&\cos{u}\partial_x+\mu\tan(\mu x)\sin{u}\partial_u,\\
e_2&=&-\sin{u}\partial_x+\mu\tan(\mu x)\cos{u}\partial_u;\\
e_0&=&\partial_u,\\
e_1&=&\cos{u}\partial_x-\mu\tanh(\mu x)\sin{u}\partial_u,\\
e_2&=&-\sin{u}\partial_x-\mu\tanh(\mu x)\cos{u}\partial_u;\\
e_0&=&\partial_u,\\
e_1&=&\cos{u}\partial_x-x^{-1}\sin{u}\partial_u,\\
e_2&=&-\sin{u}\partial_x-x^{-1}\cos{u}\partial_u.
\end{eqnarray*}
{\bf Case 2.2.}\ $W_1=0$. In this case using transformation
(\ref{r14}) with $T=t, X=x$ we can eliminate $W_4$. Inserting the
corresponding expressions for $e_1, e_2$ into (\ref{r18}) and
solving the obtained equations within the equivalence relation
${\cal E}$ yields
$$
e_0=\partial_u,\quad e_1=\cos{u}\partial_u,\quad
e_2=\sin{u}\partial_u.
$$
Note that all realizations of three-dimensional Lie algebras
obtained under Cases 2.1, 2.2 satisfy the conditions of Theorem 1.
Consequently, evolution equations invariant with respect to the
above algebras can be transformed into equations admitting QLS.

Summing up we present the full list of realizations of
three-dimensional Lie algebras, obtained for the case when $2\times
2$ matrix $C$ in (\ref{r13}) has two complex eigenvalues.
\begin{eqnarray*}
A_3^1&:&\langle\partial_u,\ \exp(\mu u)\cos{u}\partial_x
+(\mu^2+1)^{-1}(\mu\cos{u}-\sin{u})x^{-1}\exp(\mu u)\partial_u,\\
&&\quad -\exp(\mu u)\sin{u}\partial_x
-(\mu^2+1)^{-1}(\cos{u}-\mu\sin{u})x^{-1}\exp(\mu
u)\partial_u\rangle,\\
A_3^2&:& \langle \partial_u,\ \cos{u}\partial_x+\mu\tan(\mu
x)\sin{u}\partial_u, -\sin{u}\partial_x+\mu\tan(\mu
x)\cos{u}\partial_u\rangle,\\
A_3^3&:& \langle \partial_u,\ \cos{u}\partial_x-\mu\tanh(\mu
x)\sin{u}\partial_u,\ -\sin{u}\partial_x-\mu\tanh(\mu
x)\cos{u}\partial_u\rangle,\\
A_3^4&:& \langle \partial_u,\
\cos{u}\partial_x-x^{-1}\sin{u}\partial_u,\
-\sin{u}\partial_x-x^{-1}\cos{u}\partial_u\rangle,\\
A_3^5&:& \langle \partial_u,\ \cos{u}\partial_u,\
\sin{u}\partial_u\rangle.
\end{eqnarray*}
Here $\mu$ is an arbitrary real constant.

Evolution equations (\ref{r4}) invariant under the algebras
$A_3^1,\ldots,A_3^5$ can be reduced to equations admitting QLS.

\subsection {Case of diagonal canonical form with real eigenvalues}
Without loss of generality we may assume that the matrix $C$ from
(\ref{r13}) has been reduced to the diagonal matrix
$
\left(\begin{array}{cc}
\lambda_1 & 0\\[2mm]
0 & \lambda_2
\end{array}\right).
$ Since $\lambda_1, \lambda_2$ do not vanish simultaneously we may
assume that $\lambda_1\ne 0$. Re-scaling $u$ we make $\lambda_1$
equal to 1. With this choice of $C$, system (\ref{r13}) takes the
form
\begin{eqnarray*}
&&T_1=0,\quad \lambda_2T_2=0,\\
&&\xi_{1u}=\xi_1,\quad \xi_{2u}=\lambda_2\xi_2,\\
&&\eta_{1u}=\eta_1+c_1,\quad \eta_{2u}=\lambda_2\eta_2+c_2.
\end{eqnarray*}
Integrating the above system within the equivalence relation ${\cal
E}$, inserting the result into the remaining commutation relation
(\ref{r18}) and solving the equations obtained we arrive at the
following realizations of three-dimensional Lie algebras:
\begin{eqnarray*}
A_3^6&:& \langle\partial_u,\ \exp(u)\partial_x,\
(x^2+\sigma(t))\exp(-u)\partial_x + 2x\exp(-u)\partial_u\rangle,\\
A_3^7&:& \langle\partial_u,\
\exp(u)\partial_x+\epsilon\exp(u)\partial_u,\
(\sigma(t)\exp(x)\pm\exp(2x)+\mu)\exp(-u)\partial_x\\
&&\quad +(\pm\exp(2x)-\mu)\partial_u\rangle,\\
A_3^{8}&:& \langle\partial_u,\ \exp(u)\partial_x+\exp(u)\partial_u,\
(\sigma(t)\exp(-\mu x)
\pm\exp((1-\mu)x))\\
&&\quad\times\exp(\mu u)\partial_x \pm \exp((1-\mu)x))\exp(\mu
u)\partial_u\rangle,\\
A_3^{9}&:& \langle \partial_u,\ \exp(u)\partial_x,\
\sigma(t)x\exp(\mu u)\partial_x + \sigma(t)\exp(\mu
u)\partial_u\rangle,\\
A_3^{10}&:& \langle \partial_u,\ \exp(u)\partial_x,\
\sigma(t)\exp(\mu u)\partial_u\rangle,\\
A_3^{11}&:& \langle\partial_u,\ \exp(u)(\partial_x+\partial_u),\
(\sigma(t)+\exp(x))\partial_x+\exp(x)\partial_u+\epsilon\partial_t\rangle,\\
A_3^{12}&:& \langle\partial_u,\ \exp(u)(\partial_x+\partial_u),\
\sigma(t)\partial_x+\epsilon\partial_t\rangle,\\
A_3^{13}&:& \langle\partial_u,\ \exp(u)\partial_x,\
\sigma(t)(x\partial_x+\partial_u)+\epsilon\partial_t\rangle,\\
A_3^{14}&:& \langle\partial_u,\ \exp(u)\partial_x,\
\sigma(t)(\partial_x+\partial_u)+\epsilon\partial_t\rangle,\\
A_3^{15}&:& \langle\partial_u,\ \exp(u)\partial_u,\
\sigma(t)\partial_u+\epsilon\partial_t\rangle,\\
A_3^{16}&:& \langle\partial_u,\ \exp(u)\partial_u,\
\partial_x+\sigma(t)\partial_u\rangle,\\
A_3^{17}&:& \langle\partial_u,\ \exp(u)(\partial_x+\partial_u),\
\exp(u)(\sigma_1(t)\exp(-x)+\sigma_2(t))\partial_x\\
&&\quad +\sigma_1(t)\exp(u)\partial_u\rangle,\\
A_3^{18}&:& \langle\partial_u,\ \exp(u)\partial_x,\
\exp(u)(\sigma_1(t)x+\sigma_2(t))\partial_x+
\sigma_1(t)\exp(u)\partial_u\rangle,
\end{eqnarray*}
\begin{eqnarray*}
A_3^{19}&:& \langle\partial_u,\ \exp(u)\partial_u,\
x\exp(u)\partial_u\rangle,\\
A_3^{20}&:& \langle\partial_u,\ \exp(u)\partial_u,\
t\exp(u)\partial_u\rangle.
\end{eqnarray*}
Here $\sigma(t), \sigma_1(t), \sigma_2(t)$ are arbitrary real-valued
smooth functions, $\mu$ is an arbitrary real parameter, and
$\epsilon=0,1$.

By construction the coefficients of algebras $A_3^6-A_3^{20}$
satisfy the conditions of Theorem 1. Consequently, evolution
equations (\ref{r4}) invariant under these algebras can be reduced
to ones having QLS.

\subsection{Case of non-diagonal canonical form}
For the case under consideration, the matrix $C$ from (\ref{r13}) is
of the form
$ \left(\begin{array}{cc}
\lambda & 1\\[2mm]
0 & \lambda
\end{array}\right).
$
System of partial differential equations (\ref{r12}) now reads as
\begin{eqnarray*}
&&\lambda T_1 + T_2=0,\quad \lambda T_2=0,\\
&&\xi_{1u}=\lambda\xi_1+\xi_2,\quad \xi_{2u}=\lambda\xi_2,\\
&&\eta_{1u}=\lambda\eta_1+\eta_2+c_1,\quad
\eta_{2u}=\lambda\eta_2+c_2.
\end{eqnarray*}
Integrating the above system within the equivalence relation ${\cal
E}$, inserting the result into (\ref{r18}) and solving the relations
obtained we obtain eleven realizations of three-dimensional Lie
algebras:
\begin{eqnarray*}
A_3^{21}&:& \langle\partial_u,\
u\exp(u)\partial_x+x^{-1}(u+1)\exp(u)\partial_u,\
\exp(u)\partial_x+x^{-1}\exp(u)\partial_u\rangle,\\
A_3^{22}&:& \langle\partial_u,\ u\partial_x + (\mu
u^2+\sigma(t)\exp(4\mu x))\partial_u,\
\partial_x+2\mu\partial_u\rangle,\\
A_3^{23}&:& \langle\partial_u,\ u\partial_x+(\mu x + \nu u
+\sigma_1(t))\partial_u +\sigma_2(t)\partial_t,\
\partial_x\rangle,\\
A_3^{24}&:& \langle\partial_u,\ \partial_x+\mu u\tan(\mu
x)\partial_u,\ \tan(\mu x)\partial_u\rangle,\\
A_3^{25}&:& \langle\partial_u,\ \partial_x-\mu u\tanh(\mu
x)\partial_u,\ \tanh(\mu x)\partial_u\rangle,\\
A_3^{26}&:& \langle\partial_u,\ \partial_x-x^{-1}u\partial_u,\
x^{-1}\partial_u\rangle,\\
A_3^{27}&:& \langle\partial_u,\ (u^2+x)\partial_u,\
u\partial_u\rangle,\\
A_3^{28}&:& \langle\partial_u,\ (u^2+t)\partial_u,\
u\partial_u\rangle,\\
A_3^{29}&:& \langle\partial_u,\ u(\nu + 2\mu\tan(\mu t +\alpha
x))\partial_u +2\partial_t,\
(\nu + 2\mu\tan(\mu t +\alpha x))\partial_u\rangle,\\
A_3^{30}&:& \langle\partial_u,\ u(\nu + 2\mu\tanh(-\mu t +\alpha
x))\partial_u +2\partial_t,\
(\nu + 2\mu\\
&&\quad \times\tanh(-\mu t +\alpha x))\partial_u\rangle,\\
A_3^{31}&:& \langle\partial_u,\ (\mu\nu x - \mu t-2) (t-\nu
x)^{-1}u\partial_u + 2t\partial_t,\
(\mu\nu x - \mu t-2)\\
&&\quad \times(t-\nu x)^{-1}u\partial_u\rangle.
\end{eqnarray*}
Here $\sigma(t), \sigma_1(t), \sigma_2(t)$ are arbitrary real-valued
smooth functions, $\alpha, \mu, \nu$ are arbitrary real parameters.

Evolution equations (\ref{r4}) admitting symmetry algebras
$A_3^{21},\ldots,A_3^{31}$ can be reduced to equations having QLS.

\section{Some generalizations}

The technique developed in the previous sections naturally expands
to cover general systems of evolution equations
\begin{equation}
\label{r19}
\vec u_t=\vec F(t,x,\vec u,\vec u_1,\ldots,\vec u_n),\quad n \ge 2.
\end{equation}
Here $\vec F=(F^1,\ldots,F^m)$ is an arbitrary $m$-component
real-valued smooth function, $\vec u = (u^1,\ldots,u^m)\in {\mathbb
R}^m$, and $\vec u_i=(\partial^i \vec u)/(\partial x^i)$,
$i=1,\ldots,n$.

Suppose that system of evolution equations (\ref{r19}) admits
$m$-parameter Abelian symmetry group which leaves the temporal
variable, $t$, invariant. The infinitesimal generators of this group
have to be of the form
\begin{equation}
\label{r20}
Q_i=\xi_i(t,x,\vec u)\partial_x + \sum_{j=1}^m\eta_{ij}(t,x,\vec
u)\partial_{u^j}, \quad i=1,\ldots,n
\end{equation}
and what is more, the rank of the matrix composed of the
coefficients of differential operators
$\partial_x,\partial_{u^1},\ldots,\partial_{u^m}$ equals to $m$.
Given these conditions, there exists a change of variables
\begin{equation}
\label{r200}
\bar t = t,\quad \bar x = X(t,x,\vec u),\quad
\vec {\bar u} = \vec U(t,x,\vec u)
\end{equation}
that reduces operators (\ref{r20}) to canonical ones $\bar
Q_i=\partial_{{\bar u}^i}$, $i=1,\ldots,m$ (see, e.g., \cite{ovs1}).
Consequently system of evolution equations (\ref{r19}) takes the
form
\begin{equation}
\label{r21}
\vec u_t=\vec f(t,x,\vec u_1,\ldots,\vec u_n),\quad n \ge 2.
\end{equation}
Note that we dropped the bars.

Now we can apply the same trick we utilized for the case of a single
evolution equation. Namely, we differentiate (\ref{r21}) with
respect to $x$ and map $\vec u \to \vec u_x$ thus getting
\begin{equation}
\label{r22} \vec u_t=\frac{\partial \vec f}{\partial x} +
\sum_{i=1}^n\sum_{j=1}^m\,\frac{\partial \vec f} {\partial
u^j_{i-1}}u^j_i
\end{equation}
with $u^j_0\equiv u^j$.

Now if the system of evolution equations (\ref{r21}) admits the Lie
transformation group
\begin{equation}
\label{r23}
t'=T(t,\theta),\quad
x'=X(t,x,\vec u,\theta),\quad
{\vec u}'=\vec U(t,x,\vec u,\theta),
\end{equation}
where $\theta\in {\mathbb R}$ is a group parameter, then the
transformed system of equations (\ref{r23}) admits the group
\begin{eqnarray}
&t'&=T(t,\theta),\nonumber\\
&x'&=X(t,x,\vec v,\theta),\label{r24}\\
&\vec u'&=\frac {\vec U_{x} + \sum_{i=1}^m\vec U_{v^i}u^i} {X_{x} +
\sum_{i=1}^m X_{v^i}u^i}\nonumber
\end{eqnarray}
with $v^i=\partial^{-1}u^i\equiv \int\,u^idx$ and $\vec U = \vec
U(t,x,\vec v,\theta)$. Consequently, provided either of relations
\begin{equation}
\label{r25}
\frac{\partial X}{\partial v^j}\ne 0,\quad \frac{\partial}{\partial
v^j}\,\left(\frac {\vec U_{x} + \sum_{i=1}^m\,\vec U_{v^i}u^i}{X_{x}
+ \sum_{i=1}^m X_{v^i}u^i}\right)\ne 0
\end{equation}
holds for some $j,\ 1\le j \le m$, system of evolution equations
admits QLS (\ref{r24}).

Set of relations (\ref{r25}) is equivalent to the following system
of inequalities
$$
\sum_{i=1}^m\,X_{v^i}^2\ne 0,
$$
or
$$
\sum_{i=1}^m\,X_{v^i}^2= 0,\quad \sum_{i,j=1}^m\,(U^j_{xv^i})^2 +
\sum_{i,j,k=1}^m\,(U^k_{v^iv^j})^2 \ne 0.
$$
Note that all the functions involved are real-valued, so that
vanishing of the sum of squares requires that every summand vanishes
individually. Rewriting the obtained relations in terms of
coefficients of the corresponding infinitesimal operators we arrive
at the following assertion.
\begin{Theorem}
System of evolution equations (\ref{r21}) can be reduced to a system
having QLS if it admits Lie symmetry whose infinitesimal operator
$Q=\tau(t)\partial_t + \xi(t,x,\vec u)\partial_x +
\sum_{i=1}^m\eta_{i}(t,x,\vec u)\partial_{u^i}$ satisfies one of the
inequalities
\begin{equation}
\label{r26}
\sum_{i=1}^m\,\xi_{v^i}^2\ne 0,
\end{equation}
\begin{equation}
\label{r27}
\sum_{i=1}^m\,\xi_{v^i}^2= 0,\quad \sum_{i,j=1}^m\,(\eta^j_{xv^i})^2
+ \sum_{i,j,k=1}^m\,(\eta^k_{v^iv^j})^2 \ne 0.
\end{equation}
\end{Theorem}

Summing up we formulate the algorithm for classifying systems of
evolution equations (\ref{r19}) admitting QLS.
\begin{enumerate}
\item We compute the maximal Lie symmetry group ${\cal S}$ of system
of partial differential equations (\ref{r19}). \item We classify
inequivalent $m$-parameter Abelian subgroups ${\cal S}_1,\ldots{\cal
S}_p$ of the group ${\cal S}$ and select subgroups whose
infinitesimal operators are of the form (\ref{r20}).
\item For each subgroup ${\cal S}_i$ we construct change of
variables (\ref{r200}) reducing commuting infinitesimal operators,
$Q_i$, to the canonical forms $\partial_{\bar u^i}$, which leads to
system of evolution equations (\ref{r21}).
\item Since the invariance group, $\bar{\cal S}$, admitted by
(\ref{r21}) is isomorphic to ${\cal S}$, we can utilize the results
of subgroup classification of ${\cal S}$. For each of the
$m$-parameter Abelian subgroups of $\bar{\cal S}$ we check whether
their infinitesimal generators satisfy one of conditions
(\ref{r26}), (\ref{r27}) of Theorem 2. This yields the list of
systems of evolution equations that can be reduced to those having
QLS.
\item Performing the non-local change of variables $u^i\to u^i_x$,
$i=1,\ldots,m$ we obtain systems of evolution equations (\ref{r22})
admitting quasi-local symmetries (\ref{r24}).
\end{enumerate}

We intend to devote a special publication to application of this
algorithm to Schr\"odinger-type systems of partial differential
equations. Here we present an example of Galilei-invariant nonlinear
Schr\"odinger equation, which leads to the equation possessing QLS.

Consider nonlinear Schr\"odinger equation
\begin{eqnarray}
&&i\psi_t=\psi_{xx} + 2(x+i\alpha)^{-1}\psi_x - (i/2)(x+i\alpha)
\label{r28} \\
&&\quad + F(2i\alpha(x+i\alpha)\psi_x - (x-i\alpha)
(\psi-\psi^*)),\nonumber
\end{eqnarray}
where $\psi=\psi_{RE}(t,x)+i\psi_{IM}(t,x)$,
$\psi^*=\psi_{RE}(t,x)-i\psi_{IM}(t,x)$, $\alpha\ne 0$ is an
arbitrary real constant and $F$ is an arbitrary complex-valued
function. According to \cite{ren4}, Eq.(\ref{r28}) admits the Lie
algebra of the Galilei group having the following basis operators:
\begin{eqnarray}
e_1&=& \partial_t,\nonumber\\
e_2&=& \partial_{\psi} + \partial_{\psi^*},\nonumber\\
e_3&=& (x+i\alpha)^{-1}\partial_{\psi} +
(x-i\alpha)^{-1}\partial_{\psi^*},\nonumber\\
e_4&=& \partial_x-(t+(x+i\alpha)^{-1}\psi)\partial_{\psi} -
(t+(x-i\alpha)^{-1}\psi^*)\partial_{\psi^*}.\nonumber
\end{eqnarray}

Operators $e_2$, $e_3$ commute and the rank of the matrix of
coefficients of operators $\partial_t, \partial_x, \partial_{\psi},
\partial_{\psi^*}$ is equal to 2. Consequently, there is a change of
variables that reduces $e_2$, $e_3$ to canonical forms $\partial_u$,
$\partial_v$. Indeed, making the change of variables
\begin{equation}
\label{r29} u(t,x) = (1/2)(\psi+\psi^*),\quad v(t,x) =
(2i\alpha)^{-1}(x^2+\alpha^2) (\psi-\psi^*)
\end{equation}
transforms $e_1,e_2$ to become $e_1=\partial_u$, $e_2=\partial_v$.
So we can apply the above algorithm to Eq.(\ref{r28}) transformed
according to (\ref{r29}). As the coefficients of the transformed
operator $e_3$ satisfy (\ref{r27}), it leads to QLS of the system of
evolution equations of the form (\ref{r22}).

\section{Concluding remarks}

In the present paper we develop the efficient approach to
constructing evolution type partial differential equations which
admit quasi-local symmetries. It is important to emphasize that the
algorithm can be applied iteratively. Namely, if the transformed
equation possesses Lie symmetries which satisfy conditions of
Theorem 1, then it again can be transformed to a new evolution
equation admitting QLS and so on. What is more, the equation
obtained as the $N$th iteration of the algorithm admits QLS which
involves non-local variables $\partial^{-1}u,
\partial^{-2}u,\ldots,\partial^{-N}u$.

It is also of great interest to explore the case of multi-component
evolution equations. The most natural objects are the non-linear
Schr\"odinger-type equations or systems of nonlinear
reaction-diffusion equations.

There is a different approach to analyzing non-local symmetries for
some special differential equations based on the notion of potential
symmetries introduced by Bluman \cite{blu2,blu3}. It is of interest
to explore the connection between two approaches, which would open a
way to group classification of equations with potential symmetries.

Study of the above mentioned problems is in progress now and will be
reported in our future publications.

\end{document}